# Electric field tuning of magnetic states in single magnetic molecules

Yan Lu[1,‡], Yunlong Wang[1,‡], Linghan Zhu[2], Li Yang[2,*], and Li Wang[1,†]

1. Department of Physics, Nanchang University, Nanchang, 330031, People's Republic of China

2. Department of Physics and Institute of Materials Science and Engineering, Washington University, St. Louis, MO, 63130, USA

**ABSTRACT**

Single magnetic molecules may be the smallest functional magnets. An electric-field controllable spin state of magnetic molecules is of fundamental importance for applications while its realization remains challenging. To date the observed spin-electric interaction based on spin-orbit coupling or spin dipole coupling is useful to tune fine spin structures but too weak to flip the spin state. In this work, we propose a new mechanism to realize enhanced spin-electric coupling and flip the spin states by tuning the spin superexchange between local spins. Using first-principles calculations and Heisenberg Hamiltonian, we demonstrate this effect in a family of magnetic molecules, transition metallic Porphyrins. We show that their $d$-$\pi$ and $\pi$-$\pi$ spin superexchange couplings are determined by the relative energies of $d$ and $\pi$ electronic states, which are sensitive to the applied electric field. Therefore, applying electric field can tune a wide range of magnetic ground states, including ferromagnetic, ferrimagnetic, and antiferromagnetic configurations. This spin-electric coupling may provide a new approach for designing and controlling molecular spintronics.

## I. INTRODUCTION

Magnetic molecules have been regarded as the smallest functional magnets. To date a few families of magnetic molecules, including single-molecule magnets [1,2], have attracted increasing research interests because of their potential applications for high density information storage, nanospintronics, spin qubits, quantum computation, and so on [3-6]. In these applications, manipulating spin states by electric field, rather than magnetic field, is highly desirable, because electric field has the advantages of extreme space confinement, easy manipulation, high energy efficiency, and quick response [7].

Unfortunately, an isolated single spin does not couple to electric field. For spins in lattices or molecules, spin-electric coupling may be achieved through indirect processes by specific ligand fields. To date the spin-electric coupling in magnetic molecules has been achieved through spin-orbit coupling or spin-dipole moment coupling. By coordinating rare-earth ions into molecules, a strong spin-orbit coupling can induce anisotropy of the ligand field, giving rise to the spin-electric coupling [8]. For instance, this spin-orbit coupling mechanism enables the manipulation of nuclear spin resonance in Terbium-based magnetic molecules [9] by electric field, and magnetic field can enhance the dielectric permittivity of Dysprosium-based magnetic molecular crystal [10]. For magnetic molecules with permanent electric polarization, whether local or long ranged, spins can couple with the intrinsic polarization through the dipolar spin coupling [11-17]. This coupling was observed in the electron spin resonance of spin rings [18-22] and spin chains [23,24]. However, these intrinsic mechanisms are usually too weak to flip the spin configurations [5], which are crucial for information-storage applications. On the other hand, extrinsic approaches have been proposed to tune electron configurations and spin states by charge transfer between magnetic molecules and substrates while the required heterogeneous structures [25-27] are challenging for device integrations.

In this work, we propose an intrinsic, enhanced spin-electric coupling mechanism by electrically tuning the spin superexchange couplings between local spins. Using the Heisenberg model and first-principles calculations, we demonstrate this effect in a family of magnetic molecules called transition metallic Porphyrins (TM-Pcs) that possess both $d$ and $\pi$ magnetisms.

We find that the magnetic ground state is determined by the competition between *d-π* and *π-π* superexchange couplings, which are tunable by an experimentally accessible static electric field. As a result, the magnetic state can be altered between ferromagnetic (FM), ferrimagnetic (FRI), and antiferromagnetic (AFM) configurations. Finally, beyond TM-Pcs, we show that the critical electric field can be substantially reduced in other magnetic molecules, and this mechanism is also applicable for magnetic molecules with only *d* magnetism.

The article is organized in the following way. In section II, we present the computational details. In section III, we present the origins of magnetism and the exchange coupling paths in TM-Pcs. In Section IV, we show that applied electric field can tune the energy alignment of molecular orbital levels and the corresponding strength of exchange couplings, resulting in an effective spin-electric coupling to switch the magnetic configurations. In Section V, this picture is expanded to larger-size molecules, in which the critical field is significantly reduced. The results are concluded in Section VI.

**II.   Calculation Methods**

First-principles DFT implemented in the Vienna Ab initio Simulation Package (VASP) [28,29] is applied by using the Perdew-Burke-Emzemhof (PBE) [30] and Heyd-Scuseria-Ernzerhof (HSE) [31] functionals. A plane-wave basis set with a kinetic energy cutoff of 400 eV and a 1×1×1 k-point sampling are adopted for geometry optimizations and self-consistent calculations. For geometry optimizations, all atoms are fully relaxed until the residual force per atom is less than 0.01 eV/Å. The distance between magnetic molecules in neighboring supercells is larger than 12 Å to avoid spurious interactions. Except for calculations of partial density of states (PDOS), spin-orbital coupling is included in all calculations. The main data are calculated by the DFT + U functional based on the Dudarev scheme [32] with U=3 eV. Other values of U have also been tested and give qualitatively similar results. For HSE calculations, the structures of magnetic molecules are fixed according to the relaxed structures calculated by DFT+U with U=3 eV. In this work, we have considered the following transition metals, V, Cr, Fe, Co, Ni, Cu, and Zn, which cover the most popular choices in this field.

### III. Spin Configurations and Exchange Couplings in TM-Pcs

In the following, we choose TM-Pcs [33] shown in Fig. 1(a) to present this spin-electric coupling mechanism. TM-Pcs have been widely synthesized in experiments. Their excellent photophysical [34] and rich spintronic [35-39] properties have attracted significant interest. There are two origins of magnetism in TM-Pcs. First, as shown in Fig. 1 (a), each zigzag edge exhibits $\pi$ magnetism with total magnetic moments of 1 $\mu_B$ due to the open-shell character, which has been verified by recent scanning tunneling spectroscopy experiments [33]. This magnetism is like that of zigzag graphene nanoribbons [40-44]. The FM order at each edge is through the short-range direct exchange between opposite spins of the nearest carbon sites [45,46]. Secondly, a transition metal atom at the center contributes $d$ magnetism due to the partially occupied $d$ orbitals. Consequently, the overall magnetic state of TM-Pcs is determined by the coupling and competition between $\pi$ and $d$ magnetisms.

To clarify the magnetic state and couplings between local magnetic moments, we introduce $J_0$ to represent the coupling between two edge $\pi$ magnetisms and $J_1$ for that between $d$ and $\pi$ magnetisms, respectively, as denote in Fig. 1(a). To calculate the competition between local magnetic moments in such weak spin-orbit coupling systems, we describe the magnetic properties of TM-Pcs by the Heisenberg model (see section S1 of the Supplemental Material [47]): $\mathcal{H} = J_0 \boldsymbol{S}_\pi^L \cdot \boldsymbol{S}_\pi^R + J_1 \boldsymbol{S}_d \cdot \boldsymbol{S}_\pi^L + J_1 \boldsymbol{S}_d \cdot \boldsymbol{S}_\pi^R + A_x (S_d^x)^2 + A_y (S_d^y)^2$, where $\boldsymbol{S}_\pi^L$, $\boldsymbol{S}_\pi^R$, and $\boldsymbol{S}_d$ are the spins of the left $\pi$, right $\pi$, and central $d$ magnetisms, respectively. $S_d^x$ and $S_d^y$ are spin components of $\boldsymbol{S}_d$ along $x$ and $y$ directions, respectively. $A_x$ and $A_y$ are used to represent the magnetic anisotropic energy for spin along the $x$ and $y$ directions, respectively, due to the spin-orbital interaction of $3d$ ions. According to the signs and magnitudes of $J_0$ and $J_1$, TM-Pcs may exhibit five magnetic ground coupling states, *i.e.*, FM, FRI, AFM, AFM0, and FM0, as summarized in Fig. 1(b). Particularly, if the magnetic moment of the transition-metal atom is zero, $J_0 = 0$ and the molecule is in the AFM0 or FM0 state.

Next, we quantify the magnetic ground states of specific TM-Pcs. Their magnetic configurations and corresponding energies are calculated by DFT [28,29]. We plot the calculated relative total energies in Fig. 2, where that of the FM state is set to be 0 as the reference. Take V-Pc as an example, the FRI and AFM states are about 19 meV and 10 meV lower than the FM state, indicating the FRI ground state. Based on these energies, we can further extract the coupling values of $J_0$ and $J_1$, which are 0.8 meV and 6.3 meV for V-Pc, respectively. The extracted values of $J_0$, $J_1$, $A_x$, and $A_y$ for other TM-Pcs are summarized in Table I. An observation of Table I is that the magnetic ground state of most TM-Pcs is AFM or FRI.

To better understand the physics of $J_0$ and $J_1$, we have analyzed the corresponding superexchange couplings. This is also crucial for finding ways to tune them in section IV. Generally, the superexchange coupling is an indirect interaction between non-neighboring spins mediated by intermediate states [48-51]. The electron may acquire additional kinetic energy advantage by hopping between the initial and transition states, as shown in Fig. 3(a). The spin superexchange is essentially a four-step virtual hopping process in our studied TM-Pcs, and the coupling $J$ can be evaluated by the perturbation theory [48-51].

$$J = \frac{\langle i|H_{ip}|p\rangle}{E_p - E_i} \cdot \frac{\langle p|H_{pt}|t\rangle}{E_t - E_p} \cdot \frac{\langle t|H_{tp}|p\rangle}{E_p - E_t} \cdot \frac{\langle p|H_{pi}|i\rangle}{E_i - E_p} \quad (1)$$

where $|i\rangle$, $|t\rangle$, and $|p\rangle$ are the initial, transition, and intermediate $p$ orbitals marked in Fig. 3(a), respectively. $E_i$, $E_t$, and $E_p$ are the corresponding orbital energies. $H_{ip}$ and $H_{pt}$ are the hopping Hamiltonians. Within practical electric field, the changes of molecular orbitals and hopping Hamiltonians are negligible. Thus, the electric field effect on $J$ is mainly from the change of relative orbital energies, the denominator in Eq. (1). More specifically, we find that the energy of the intermediate $p$ orbital is close to that of the occupied state $i$ in our studied TM-Pcs, which will be seen from the calculated electronic structures later. Thus, the leading order of the superexchange coupling is reduced to be

$$J \propto \frac{1}{(E_t - E_i)^2}, \quad (2)$$

in which $J$ is inversely proportional to the square of the energy difference between the initial and transition states.

Following this picture, we employ DFT to calculate the PDOS of TM-Pcs to determine the positions of $E_i$ and $E_t$. Fig. 3(b) shows the example of V-Pc. The orange and wine-red curves are the PDOS of the spin-polarized $\pi$ orbitals, and the green ones are from the $d$ orbitals. With this PDOS, we can analyze the superexchange coupling paths for $J_0$ and $J_1$. First, we focus on $J_0$, which is the coupling between two edge $\pi$ orbitals. Figures 3(c) and (d) schematically plot the AFM and FM coupling paths, respectively. For the AFM superexchange paths in Fig. 3 (c), the energy difference between initial and transition states is the energy gap between the highest occupied molecular orbital (HOMO) $\pi$ and lowest unoccupied molecular orbital (LUMO) $\pi$ orbitals, and it is around 0.4 eV in Fig.3 (b). Due to this small energy barrier, the AFM superexchange strength of $J_0$ is strong based on Eq. (2). For the FM coupling of $J_0$ shown in Fig. 3(d), the superexchange path between HOMO and LUMO $\pi$ states is blocked because of the Pauli's exclusion principle. Thus, most TM-Pcs do not exhibit an overall FM configuration between edge $\pi$ magnetism.

The superexchange couplings of $J_1$ between $\pi$ and $d$ states can be understood in the similar way. In Figs. 3(e) and 3(f), we plot the dominant AFM and FM superexchange couplings of $J_1$ for V-Pc, respectively. In Fig. 3(e), the initial and transition states are HOMO ($d$) and LUMO ($\pi$) for the AFM superexchange. The energy difference is about 0.4 eV, as read from the dashed square in Fig. 3 (b). On the other hand, Fig. 3(f) shows that the initial and transition states are HOMO ($\pi$) and LUMO ($d$) for the FM superexchange, and the energy barrier is about 1.9 eV in Fig. 3(b). This indicates a stronger AFM coupling than that of FM coupling. Combined with the above conclusion of the preferred AFM coupling between edge $\pi$ magnetisms, this explains why the overall FRI state is more stable than the FM state for V-Pc shown in Fig. 2.

### IV.    Enhanced Spin-Electric Couplings in TM-Pcs

The above discussions reveal that the energy difference between the initial and transition states is important to determine the strength of the corresponding superexchange coupling. Therefore, by tuning the energy difference between $d$ and $\pi$ orbitals, we may control the strength of the corresponding superexchange coupling and realize the enhanced spin-electric coupling of TM-Pcs.

Thanks to the spatial separation between $d$ and $\pi$ local magnetic moments, an effective way of tuning their energy differences is to apply a static electric field along the $\pi$-$d$-$\pi$ direction. For V-Pc, the superexchange for the FRI and AFM configurations are shown in Figs. 4(a) and 4(b), respectively. The superexchange paths for the FRI state include the paths A and C, as shown in Fig. 4(a). The superexchange paths for the AFM state include the paths B1, B2, and C, as shown in Fig. 4(b). Importantly, the path C in these two spin states is nearly the same. Therefore, the energy different between FRI and AFM states is essentially determined by the superexchange strengths of path A in the FRI state, and the paths B1 and B2 in the AFM state.

As shown in Fig. 4(c), the energy gap between the $d$ and right $\pi$ edge states is increased by electric field. According to Eq. (2), the strength of superexchange path A ($d$-$\pi$) is decreased by electric field. On the other hand, this electric field decreases the energy gap of the path B1 and increases the energy gap of the path B2, as shown in Fig. 4(d). According to Eq. (2), the exchange is dominated by the path B1 because of its smaller energy difference, and the AFM coupling between two $\pi$ edge states is enhanced. Therefore, under electric field, the FRI state in Fig. 4(c) becomes less stable, while the AFM state in Fig. 4(d) is more stable. Ultimately, the AFM state may be the magnetic ground state if the electric field is strong enough.

These pictures are confirmed by DFT calculations. We have calculated the energy changes after applying electric field along the $\pi$-$d$-$\pi$ direction. For V-Pc shown in Fig. 5(a), the ground state is FRI without electric field, and the energy difference between FRI and AFM is decreased under electric field. After a critical electric field about 0.17 V/Å, the ground state becomes AFM, which confirms our predictions. Interestingly, in addition to the change of magnetic configurations, the total magnetic moment of V-Pc is changed from 1 $\mu_B$ to 3 $\mu_B$. This brings a new degree of freedom for information storage applications. For another type of TM-Pc, *i.e.*, Co-Pc, the critical electric field is about 0.12 V/Å, where an FM to AFM transition occurs, with the total magnetic moment changed from 3 $\mu_B$ to 1 $\mu_B$. It is worth mentioning that this amplitude of electric field is well within current experimental capability [52,53], making our predictions useful for realizations. The DFT calculations also show that the magnetic anisotropic energy parameters, $A_x$ and $A_y$, are nearly constant under the electric field within a reasonable range. (See Section S7 of the Supplementary Materials).

The model of Eq. (2) can quantitatively estimate the field-tunable superexchange strengths of paths A and B. According to Eq. (2), the key is to find the energy difference between initial and transition states. The DFT-calculated PDOS of V-Pc around the Fermi level is replotted in Figs. 6(a) and (b) for FRI and AFM states, respectively. In Fig. 6(b), $\Delta_1$ is the energy difference between initial and transition states of the path B1 in Fig. 4(b). $\Delta_2$ and $\Delta_3$ are the energy differences between initial and transition states of the path B2, which has been split into two sub paths due to the hybridization between the occupied $\pi$ and $d$ orbitals. According to Eq. (2), for the AFM state, we have

$$J_{S0} \propto \left(\frac{1}{\Delta_1^2} + \frac{\rho_2}{\Delta_2^2} + \frac{\rho_3}{\Delta_3^2}\right) \tag{3}$$

where the first term is the strength of the superexchange path B1, the second and third terms are strengths of the superexchange path B2 shown in Fig. 4(b). The corresponding weights $\rho_2$ and $\rho_3$ can be obtained by the DFT-calculated split peak heights shown in Fig. 6(b). The change of $\Delta_1$, $\Delta_2$, and $\Delta_3$ with electric field are plotted in Fig. 6(c). According to Eq. (3), the change of $J_{S0}$ is plotted in Fig. 6(d), which is nearly identical to the change of DFT-calculated $J_0$ of the AFM state. In Fig. 6 (e), for the $d$-$\pi$ superexchange coupling, the Eq. (2)-calculated change of $J_{S1}$ also reproduce the DFT-calculated $J_1$. A slight deviation indicates that the change of the $d$-$\pi$ coupling under electric field is more complicated due to the existence of multiple $d$ orbitals and subsequent extra $d$-$\pi$ superexchange sub-paths.

With obtained $J_{S0}$ and $J_{S1}$, we can calculate the total energy of the AFM and FRI magnetic states of V-Pc, as shown in Fig. 6(f). The critical electric field is about 0.165 V/Å for a transition from the FRI to AFM configuration. This estimation excellently agrees with our DFT calculations (0.17 V/Å) shown in Fig. 5(a).

It is known that DFT calculations usually underestimate energy gaps of materials or give unreliable energies of correlated states. Because the magnetic ground state is sensitively to the energy levels of $d$ and $\pi$ states, the DFT-calculated ground states as well as the electric field tuned magnetic states may not be accurate. To answer this concern, we have employed the DFT+$U$ and HSE [31] to improve the calculations of energy levels and energy gaps to check the electric-field tuned magnetic states. First, we have checked these results with different values of Hubbard $U$ in DFT+$U$ calculations, and the general results remains the same, although the critical field can be changed by the values of $U$. (See Section S2 of the Supplementary Materials).

Meanwhile, we find HSE correct the band gap and does affect the magnetic ground states of TM-Pcs. For V-Pc, the ground state is FRI in DFT calculation. However, the HSE ground state is FM, whose energy is about 38 meV and 45 meV lower than those of FRI and AFM states, respectively. Based on Eq. (2), the AFM interaction strengths of both $J_0$ and $J_1$ will be decreased significantly in HSE calculations. Such decreases can be quantitatively understood from the change of the energy levels by HSE calculations. The energy gap between HOMO ($\pi$) and LUMO ($\pi$) states is increased from 0.4 to 1.3 eV in HSE calculations. Therefore, the coupling strengths will decrease substantially for superexchange paths between the initial and final $\pi$ orbitals, such as the AFM interactions of both $J_0$ and $J_1$ shown in Figs. 3(c) and 3(e). (See hybrid functional details in Section S3 of the Supplementary Materials)

Despite the numerical values from different approaches, the general picture of our proposed electric-spin coupling remains valid: the energy difference between FM and AFM states can be decreased as the energy levels between $d$ and $\pi$ states are changed under electric field. For example, the ground-state of V-Pc is switched from the FM state to the AFM state under an electric field of 0.65 V/Å. The larger critical field is from the larger gap and corresponding lower energy of the FM state (~45 meV) in HSE simulations. In Sections S4 of the supplementary materials, we have provided potential tip and gate materials for implementing the electric field.

## V. Enhanced Spin-Electric Coupling in Larger-Size Molecules

Practically a smaller critical field is desired for applications and experiments. To reduce the critical field, we propose TM-2Pcs, the larger-size molecules, as shown in Fig. 7(a), which have been synthesized recently by the bottom-up method [54]. In these TM-2Pcs, there also exists $d$ and $\pi$ magnetisms. The main difference is that the $d$-$\pi$ coupling on the left and right sides are different. Importantly, the HSE simulations find that the magnetic states of TM-2Pcs can be tuned by smaller electric field, under the same physics mechanism discussed above. Take Cr-2Pc as an example, the ground state is AFM, while the FM one becomes the ground state under a smaller critical electric field of 0.015 V/Å, as denote by the wine-red dashed line in Fig. 7(c). The field sign is defined as positive from Porphyrin with transition metal to Porphyrin without transition metal. For Fe-2Pc, the critical electric field is about -0.05 V/Å, where an FM/AFM transition occurs. With this FM to AFM transition, we also observe that the total magnetic moments of Fe-2Pc and Cr-2Pc are changed from 4 $\mu_B$ to 2 $\mu_B$ and from 6 $\mu_B$ to 4 $\mu_B$, respectively. In Section S5 of the Supplementary Materials, we show that details of the calculation of TM-2Pcs and this critical field can be further reduced in larger molecules, such as TM-3Pcs.

Finally, this proposed competition mechanism for realizing spin-electric field is a general picture, and it can be applied to more types of molecules, in which local spins are not necessary to be both $d$ and $\pi$ magnetisms at the same time. We have calculated an example of a TM1-TM2-2Pcs, which has two transition metals (Cr and Fe) with $d$ magnetism, as shown in Fig. 7(b). Our first-principles calculations show that the magnetic coupling between two transition metals can be changed between AFM and FM, as shown in Fig. 7(d). This electric-spin effect is from a competition between the superexchange interaction between Cr and Fe and the direct exchange between nearest sites. (see detailed discussion in section S6 of the Supplemental Material [47])

## VI. CONCLUSIONS

In conclusion, using first-principles calculations and Heisenberg model, we show that TM-Pcs can possess both $d$ and $\pi$ magnetisms. We show that their superexchange strengths can be tuned by applying electric field. Therefore, the magnetic ground states of TM-Pcs can be switched by electric field. Furthermore, beyond TM-Pcs, we show that the critical electric field can be greatly reduced in TM-2Pcs and other magnetic molecules, and this mechanism is also applicable for magnetic molecules only with $d$ magnetism. The enhanced spin-electric coupling opens a new avenue to manipulate spin configurations of magnetic molecules.

ACKNOWLEDGMENTS

Y.L. and Y.W. are supported by the National Natural Science Foundation of China (Grants No. 12164026 and No. 11504158). L.Z. is supported by the National Science Foundation (NSF) Grant No. DMR-2124934, and L.Y. is supported by the NSF Grant No. DMR-2118779. L.W. acknowledges the support from Jiangxi Provincial Innovation Talents of Science and Technology under Grant No. 20165BCB18003.

*lyang@physics.wustl.edu

†liwang@ncu.edu.cn

‡These authors contributed equally to the paper.

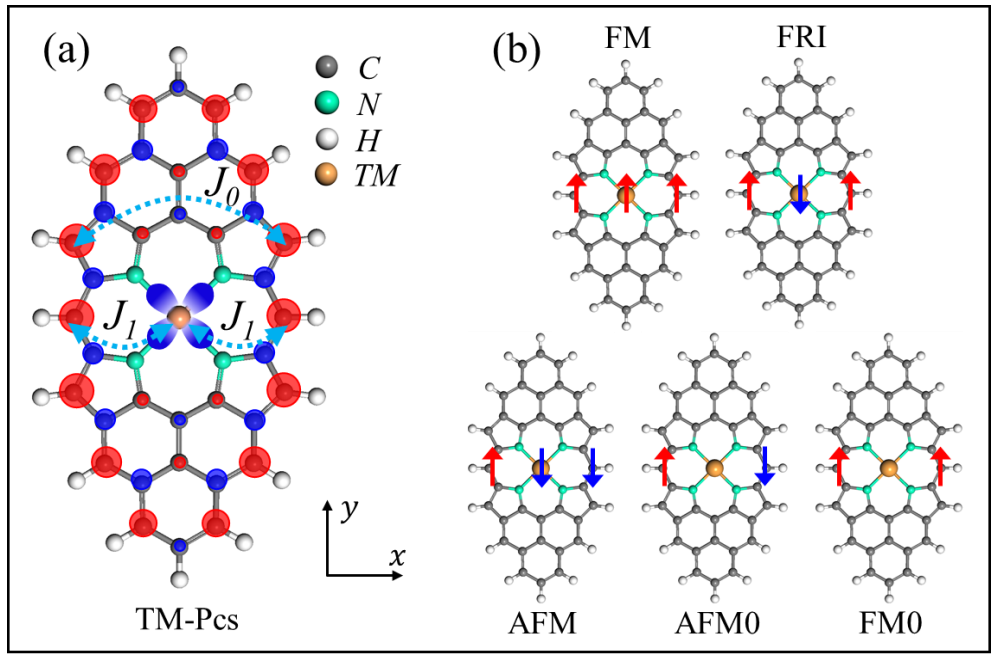

Fig. 1. Structure and magnetic states of TM-Pcs. (a) Top view of TM-Pcs. Spin up and spin down electrons are represented by red and blue colors, respectively. $J_0$ and $J_1$ denote the $\pi$-$\pi$ and $d$-$\pi$ magnetic coupling. (b) Possible five magnetic states of TM-Pcs. The red and blue arrows in (b) denote spin up and spin down, respectively.

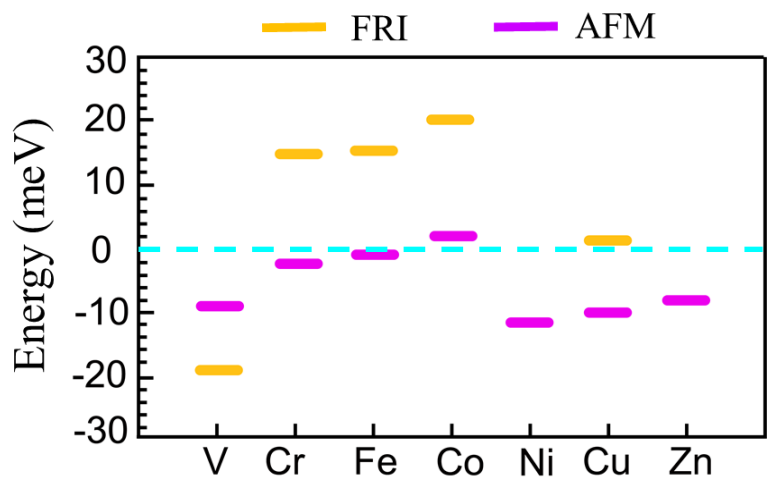

Fig. 2. DFT-calculated relative energies of possible magnetic states for TM-Pcs, where the energy of the FM state is set to be 0.

TABLE. I Extracted $m_d$, $J_0$, $J_1$, $A_x$, and $A_y$ according to the DFT-calculated energies. The units are $\mu_B$ for $m_d$, and meV for $J_0$, $J_1$, $A_x$, and $A_y$.

|       | V     | Cr    | Fe    | Co    | Ni   | Cu   | Zn   |
|-------|-------|-------|-------|-------|------|------|------|
| $m_d$ | 3     | 4     | 4     | 1     | 0    | 1    | 0    |
| $J_0$ | 0.8   | 18.4  | 14.8  | 15.2  | 24.4 | 21.2 | 16.4 |
| $J_1$ | 6.3   | -4.0  | -7.0  | -20.0 | 0    | -1.2 | 0    |
| $A_x$ | -0.07 | 0.49  | -3.4  | -0.58 | 0    | 0.11 | 0    |
| $A_y$ | 0.35  | 0.44  | -3.7  | 0.67  | 0    | 0.09 | 0    |

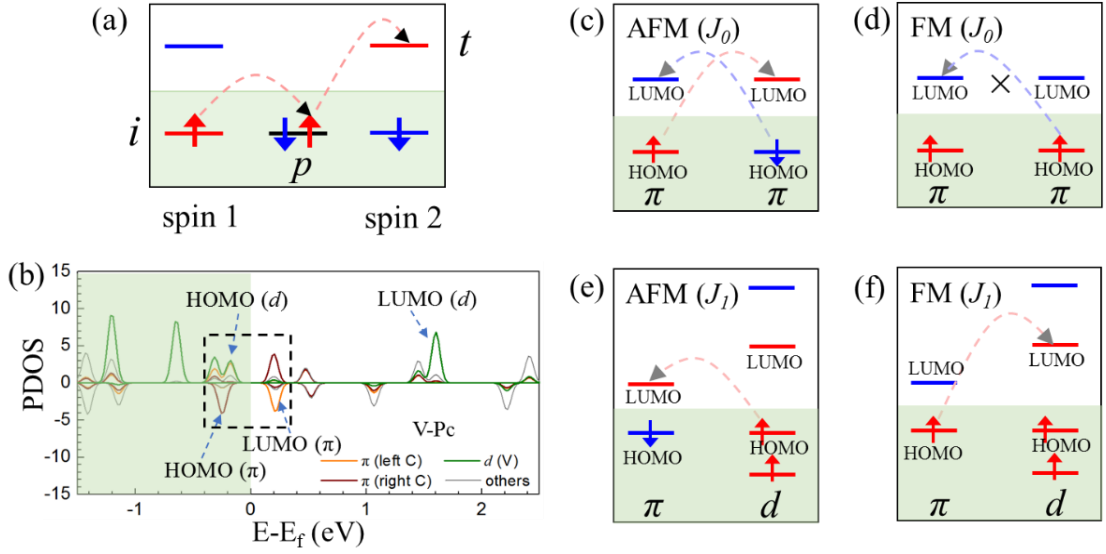

Fig. 3. PDOS and superexchange coupling paths of $J_0$ and $J_1$ for V-Pc calculated by DFT. (a) Spin superexchange between two local spins through intermediate states. The electron may acquire additional kinetic energy advantage by hopping between the initial ($i$) and transition ($t$) states. The intermediate states are coming from the $p$ orbital of C and N atoms. (b) PDOS of V-Pc calculated by DFT+$U$, in which a dashed square is used to enclose the electronic states around band edges. (c) The AFM superexchange coupling paths of $J_0$. The initial and transition states of these superexchange paths are HOMO ($\pi$) and LUMO ($\pi$), respectively. (d) The FM coupling paths of $J_0$. There are no superexchange paths between HOMO ($\pi$) and LUMO ($\pi$) due to Pauli's exclusion principle. (e) The dominated AFM superexchange coupling paths of $J_1$ for V-Pc. The initial and transition states are HOMO ($d$) and LUMO ($\pi$). (f) The dominated FM superexchange coupling paths of $J_1$ for V-Pc. The initial and transition states are HOMO ($\pi$) and LUMO ($d$). The intermediate states are not shown in (c-f). The red and blue colors denote states with spin up and spin down, respectively.

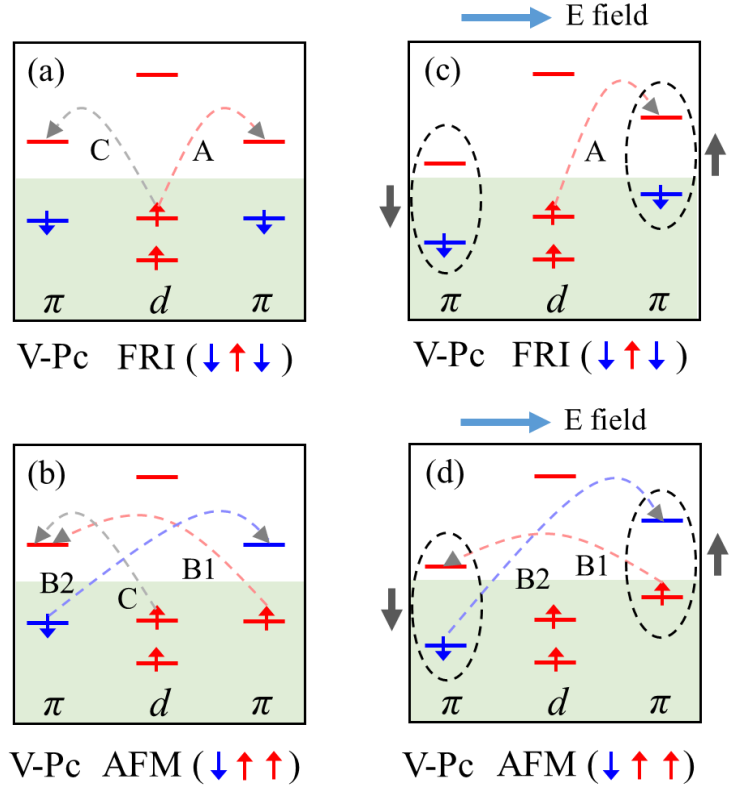

Fig. 4. Change of superexchange couplings for V-Pc under electric field. (a) Couplings at the FRI state. Path A represents the $d$-$\pi$ superexchange on the right side, path C represents the $d$-$\pi$ superexchange on the left side. (b) Couplings at the AFM state. Path C represents the $d$-$\pi$ superexchange on the left side, whose strength equals to that of the path C in (a). Paths B1 and B2 represent the $\pi$-$\pi$ superexchange couplings between two outer edges with spin-down and spin-up, respectively. (c) Couplings at the FRI state under electric field. The energy gap for path A is increased by electric field, and its strength is decreased. (d) Couplings at the AFM state under electric field. The energy gap for path B1 is decreased by field, and its strength is increased.

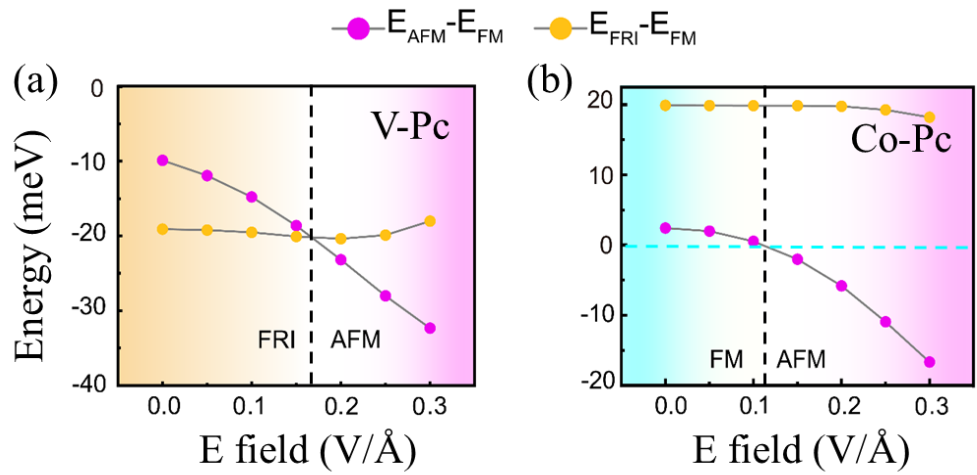

Fig. 5. Change of magnetic ground states for V-Pc and Co-Pc under electric field calculated by DFT. (a) Energy difference between FRI and FM, and between AFM and FM of V-Pc as a function of electric field calculated by DFT. A magnetic transition from FRI to AFM after a critical electric field of 0.17 V/Å. (b) Similar to (a), but for Co-Pc. A magnetic transition from FM to AFM after a critical electric field of 0.12 V/Å.

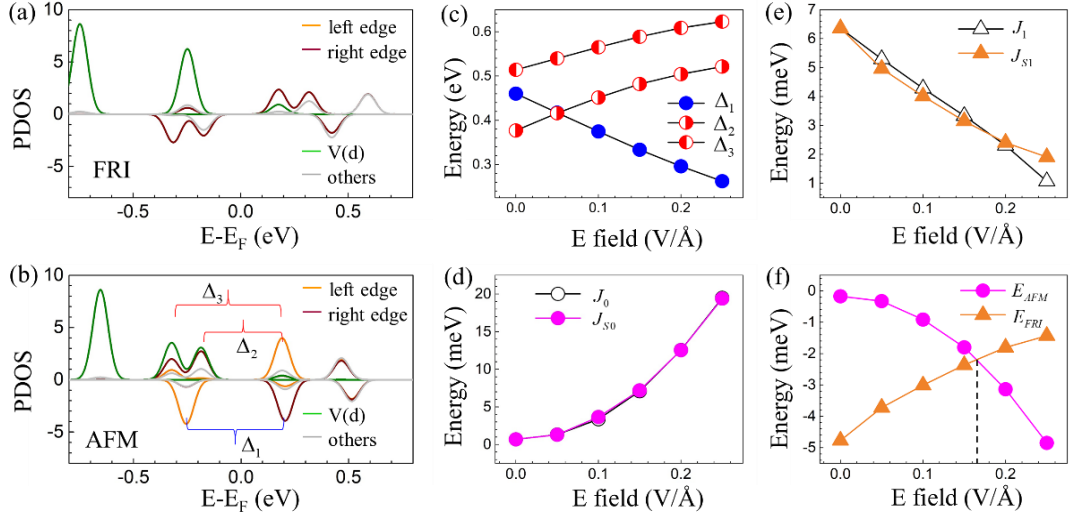

Fig. 6. Estimation of the change of superexchange strength of V-Pc according to Eq. (2). (a) PDOS of V-Pc at the FRI state. (b) PDOS of V-Pc at the AFM state. (c) Change of three energy differences as denoted in (b) under electric field. (d) Change of the π-π superexchange strength according to Eq. (2) ($J_{S0}$) and that obtained by DFT ($J_0$). (e) Change of d-π superexchange strength according to Eq. (2) ($J_{S1}$) and that obtained by DFT ($J_1$). (f) Change of total energies of AFM and FRI states of V-Pc under electric field according to Eq. (2). The critical electric field of 0.165 V/Å.

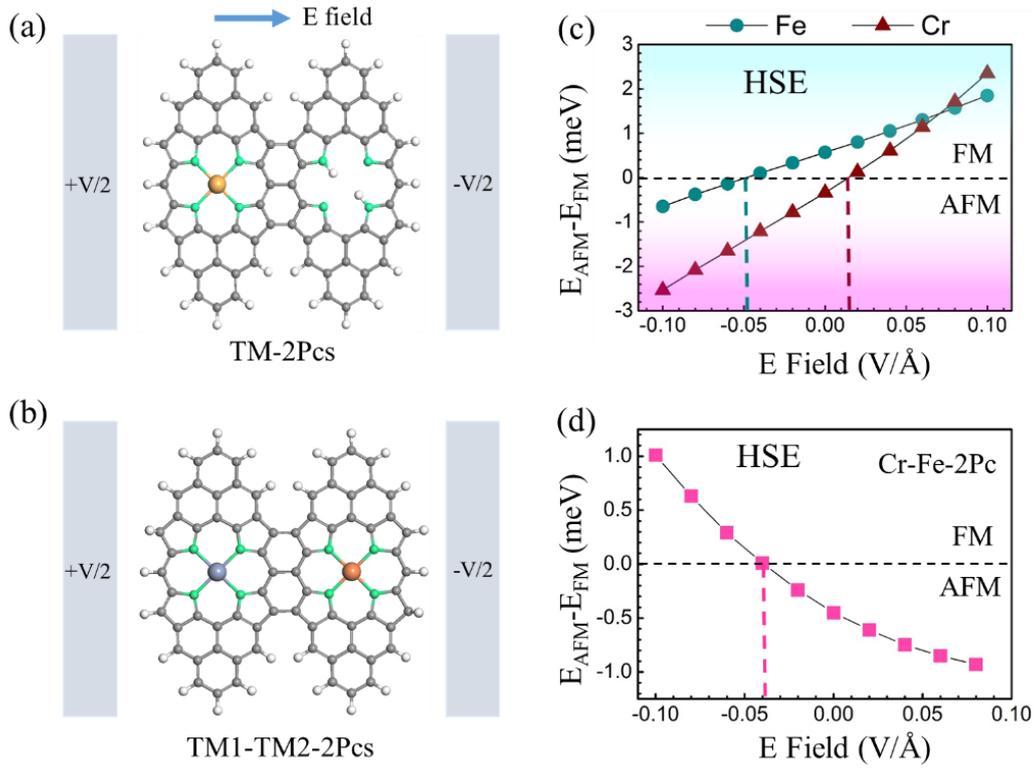

Fig. 7. Structures and the change of ground states of TM-2Pcs and TM1-TM2-2Pcs under electric field calculated by HSE. Diagram of (a) TM-2Pcs and (b) TM1-TM2-2Pcs under electric field. Energy differences between AFM and FM for (c) Fe-2Pc and Cr-2Pc, and (d) Cr-Fe-2Pc as a function of electric field calculated by HSE.